\documentclass[12pt,secnumarabic,showpacs,preprintnumbers,amsmath,amssymb]{revtex4}

\usepackage{fmtcount}
\usepackage{graphicx}
\usepackage{dcolumn}
\usepackage{bm}
\usepackage{slashed}

\begin{document}

\title{\bf Lie $3-$algebra and super-affinization of split-octonions}

\author{ Hector L. Carri\'{o}n}
\email{hectors@ect.ufrn.br}
\affiliation{Escola de Ci\^{e}ncias e Tecnologia, Universidade Federal
  do Rio Grande do Norte,\\ Av. Hermes da Fonseca 1111, 59014-615,
  Natal, RN, Brazil.}

\author{ Sergio Giardino}
 \email{jardino@fma.if.usp.br}
\affiliation{ Instituto de F\'{i}sica, Universidade de S\~{a}o Paulo, CP 66318, 05315-970,
  S\~{a}o Paulo, SP, Brazil.}

\begin{abstract}
\noindent 
The purpose of this study is to extend the concept of a generalized
Lie $3-$ algebra, known to the divisional algebra of the
octonions $\mathbb{O}$, to split-octonions $\mathbb{SO}$, which is non-divisional. This is achieved through the unification of the product of
both of the algebras in a single operation. Accordingly,
a notational device is introduced to unify the product
of both algebras.  We verify that $\mathbb{SO}$ is a
Malcev algebra and we recalculate known relations for the structure
constants in terms of the introduced structure tensor. Finally we 
construct the manifestly super-symmetric $\mathcal{N}=1\;\mathbb{SO}$
affine super-algebra. An
application of the split Lie $3-$algebra for a Bagger and Lambert gauge
theory is also discussed.
\end{abstract}

\maketitle

\section{Introduction}
The gauge-string correspondence \cite{Witten:1998qj} is one of the
most influential ideas in contemporary theoretical physics,
particularly as a model for fundamental interactions and as a
candidate for a unifying theory. The first specific example of such a
correspondence, developed by Maldacena \cite{Maldacena:1997re},
has established a duality between the {\bf IIB} string theory in the
$AdS_5\times S^5$ curved background and the $\mathcal{N}=4$,
$4-$dimensional super Yang-Mills gauge theory (SYM). Another example
is the ABJM model \cite{Aharony:2008ug}, which relates string
theory in the $AdS_4\times\mathbb{CP}^3$ background to the $3-$dimensional
$\mathcal{N}=6$ Chern-Simons theory.

Recently, Bagger and Lambert
\cite{Bagger:2006sk,Bagger:2007jr,Bagger:2007vi} and also Gustavsson
\cite{Gustavsson:2007vu,Gustavsson:2008dy} have proposed a
super-symmetric action for a stack of $M-$branes which is a
Chern-Simons model with a $3-$algebra based gauge symmetry written as 
\begin{align}
&\delta X^I=i\,\overline{\epsilon}\Gamma^I\Psi\\
&\delta\Psi=\partial_\mu X^I\Gamma^\mu \Gamma^I\epsilon+i\kappa[X^I,X^J,X^K]\Gamma^{IJK}\epsilon.
\end{align}
In these transformations, we observe the $3-$algebra that appears in
the totally anti-symmetric bracket $[X^I,X^J,X^K]$. See also
\cite{Krishnan:2008zm} for an example of application. From the $3-$algebra one can define 
a metric which can be either positive definite or not. The original
studies involve Euclidean positive definite metrics, but non-positive
Lorentzian metrics were soon introduced
\cite{deMedeiros:2008bf,DeMedeiros:2008zm,Gomis:2008uv}
along with a non-anti-symmetric triple product \cite{Bagger:2008se} gauge
theory. As the ABJM model is also a Chern-Simons theory, the
introduction of a triple product in this context has also been studied
\cite{SheikhJabbari:2008wq,Terashima:2009fy}. Further research
involving triple algebras has been carried out in string theory
\cite{Lee:2009ue} and in a more mathematical sense as a graded
super-algebra \cite{Palmkvist:2009qq}. 

These $3-$algebras are not necessarily associative. Non-associative
algebras can be divisional algebras, 
where if $a,b$ are in a divisional algebra, $\mathbb{A}$, $ab=0$
implies that either $a=0$ or $b=0$. Kugo and Townsend
\cite{Kugo:1982bn} have shown that
Lorentzian space-time spinors are associated to $\mathbb{A}$, based on
the fact that $\overline{SO}(n+1,1)=Sl(2,\mathbb{A})$ to the four normed division
algebras,  and $n$ is
related to the dimension of the field $\mathbb{A}$. Some research on the role of exceptional symmetries in
physics has been carried out involving these non-associative algebras
\cite{Ramond:2003xt,Witten:2002ei,Toppan:2003dt,Duff:2006rf}.
Non-divisional algebras, where $ab=0$ does not mean that either
$a$ or $b$ is null, can also be constructed and they have applications
in, for example, $M-$theory \cite{Kuznetsova:2006ws}. 

Finally, a non-associative algebra can be a Malcev
algebra, an extension of a Lie algebra with non-zero Jacobian. An
application of this fact is the
construction of super-affine Lie algebras
\cite{Carrion:2001pe,Lin:2008qp}. This kind of construction is related
to the current algebra, a subject which has several applications in physics, like the
construction of a super-conformal manifestly $\mathcal{N}=8$ super-symmetric hamiltonian
\cite{Carrion:2001hr}, a non-linear sigma model \cite{Benichou:2010rk},
and a Lax pair for string theory in the Green and Schwarz formalism
\cite{Linch:2008nt}. In this article we give a novel construction of
an affine super-algebra for the split-octonion case.

This article is organized as follows: in section \ref{s-o_algebra} the
octonion and split-octonion algebras are discussed and a new
formalism is introduced, and in section \ref{super-affin} a super-affinization of
the $\mathbb{SO}$ algebra is constructed. The notion of a generalized
split Lie $3-$algebra is introduced in \ref{gsl3} and in section \ref{gauge} a possible
gauge theory constructed from the realization of this algebra is discussed.

\section{The split-octonion algebra\label{s-o_algebra}}
There are four division algebras: the real, complex, quaternionic and octonionic
numbers \cite{Baez:2001dm}. From the complex,
quaternionic and octonionic numbers, non-divisional, or split-algebras
can be constructed. This means that, in these split-algebras,
$ab=0$ does not mean that either  $a=0$ or $b=0$. One example of this
is the split-octonion algebra \cite{Foot:1988at}.
To discuss the split-octonion algebra $\mathbb{SO}$ in conjunction
with the octonion algebra $\mathbb{O}$ some
notations are introduced. First, we assign $\mathbb{(S)O}$ to everything which is valid
  for both the algebras. Let us start
with a description of the octonion algebra. A common notation to an
octonion $a$ is written as $a=A_0 E_0 + A_\mu E_\mu$, where $E_0$
is the real component and $E_\mu$ are the imaginary components of the
octonion. The Greek indices have rank $\mu={1,\dots,7}$ and $A_{i=0,\dots,7}$
are real numbers. The $\mu$ index is of course summed in this
notation. The products of the base elements obey the following
multiplication table
\begin{table}[h]
\caption{\label{T1} Octonion Multiplication Table}
\begin{ruledtabular}
\begin{tabular}{ccccccccc}
              & $0$ & $1$  & $2$ & $3$ & $4$ & $5$ & $6$ & $7$ \\ 
\hline
0 $\;$ \vline & 0 & 1 & 2 & 3 & 4 & 5 & 6 & 7   \\ 
1 $\;$ \vline & 1 &-0 & 4 & 7 &-2 & 6 &-5 &-3 \\ 
2 $\;$ \vline & 2 &-4 &-0 & 5 & 1 &-3 & 7 &-6 \\ 
3 $\;$ \vline & 3 &-7 &-5 &-0 & 6 & 2 &-4 & 1 \\ 
4 $\;$ \vline & 4 & 2 &-1 &-6 &-0 & 7 & 3 &-5 \\ 
5 $\;$ \vline & 5 &-6 & 3 &-2 &-7 &-0 & 1 & 4 \\ 
6 $\;$ \vline & 6 & 5 &-7 & 4 &-3 &-1 &-0 & 2 \\ 
7 $\;$ \vline & 7 & 3 & 6 &-1 & 5 &-4 &-2 &-0 \\ 
\end{tabular}
\end{ruledtabular}
\footnotetext{ $i=E_i,\, i=0,1,\dots , 7 $ }
\end{table} 
\noindent where $E_i=i$ was used. This table is, of course, a choice, as
by changing the numbers
corresponding to each element we change the table. For this choice, we
can summarize the product of octonion base elements in the following
set of laws: 
\begin{eqnarray}
&&E_\mu E_\nu = -\delta_{\mu\nu}E_0+c_{\mu\nu\kappa}E_\kappa,\qquad
E_0E_\mu=E_\mu E_0=E_\mu,\nonumber\\
&& E_0^2=E_0,\qquad\mbox{and, to} \;\mu\neq\nu,\qquad \;E_\mu E_\nu=-E_\nu E_\mu.\label{prod_1}
\end{eqnarray}
%
%
\noindent The totally anti-symmetric tensor $c_{\mu\nu\kappa}$ gives
the structure constants of the algebra. To obtain an
explicit realization we consider that the octonion basis also obeys
\begin{equation}
E_\mu E_{\mu+1}=E_{\mu+3}\qquad\mbox{and}\qquad
E_{\mu+7}=E_\mu.
\end{equation} 
\noindent Finally, using  (\ref{prod_1}) we get the set of non-zero
structure constants
\begin{equation}
c_{124}=c_{235}=c_{346}=c_{457}=c_{561}=c_{672}=c_{713}=1.\label{str_c}
\end{equation}
%
%
\noindent Octonion algebra has several sub-algebras. There are seven associative
and non-commutative sub-algebras, isomorphic to the quaternion algebra, 
$\mathbb{H} \sim \{E_0,E_\mu,E_{\mu+1},E_{\mu+3}\}$, seven associative and
  commutative sub-algebras, isomorphic to the complex number system
  $\mathbb{C}\sim\{E_0,E_\mu\}$ and one real associative and commutative
  algebra isomorphic to the real number system, so that
  $\mathbb{R}\sim\{E_0\}$. 

The structure constants of octonion algebra (\ref{str_c})
constitute a totally anti-symmetric tensor, but in split-octonion
algebra, the structure constants no longer have this property. To describe
the situation, let us write the multiplication table 
\begin{table}[h]
\caption{\label{T2}Split-octonion Multiplication Table}
\begin{ruledtabular}
\begin{tabular}{ccccccccc}
              & $0$ & $1$  & $2$ & $3$ & $4$ & $5$ & $6$ & $7$ \\ 
\hline
0 $\;$ \vline & 0 & 1 & 2 & 3 & 4 & 5 & 6 & 7   \\ 
1 $\;$ \vline & 1 & 0 & 4 &-7 & 2 &-6 &-5 &-3 \\ 
2 $\;$ \vline & 2 &-4 &-0 & 5 & 1 &-3 & 7 &-6 \\ 
3 $\;$ \vline & 3 & 7 &-5 & 0 &-6 &-2 &-4 & 1 \\ 
4 $\;$ \vline & 4 &-2 &-1 & 6 & 0 &-7 & 3 &-5 \\ 
5 $\;$ \vline & 5 & 6 & 3 & 2 & 7 & 0 & 1 & 4 \\ 
6 $\;$ \vline & 6 & 5 &-7 & 4 &-3 &-1 &-0 & 2 \\ 
7 $\;$ \vline & 7 & 3 & 6 &-1 & 5 &-4 &-2 &-0 \\ 
\end{tabular}
\end{ruledtabular}
\footnotetext{ $i=E_i,\, i=0,1,\dots , 7 $ }
\end{table} 

\noindent where $E_{i=0,\dots,7}=i$, also. The split-octonion algebra
$\mathbb{SO}$ has the same number of sub-algebras as the octonion
case, but the majority of them are no longer division algebras. The four
component sub-algebras are isomorphic to the split-quaternion algebra,
so that  $\mathbb{SH} \sim \{E_0,E_\mu,E_{\mu+1},E_{\mu+3}\}$ with
the exception of $\mathbb{H}\sim\{E_0,E_2,E_6,E_7\}$. The sub-algebras
of order two of the type $\{E_0,E_\mu\}$ are isomorphic either to
complex or split-complex $\mathbb{SC}$ number
systems, so that $\mu=\{1,3,4,5\}$ generates split-complex
isomorphic algebras and  $\mu=\{2,6,7\}$ generates complex
isomorphic algebras. The split cases do not generate totally
anti-symmetric structure constants. As an example, in Table\ref{T2} we
see that $c_{124}=c_{142}=1$.  We do not know a multiplication law like
(\ref{prod_1}) for the split-octonion case. A proposal has appeared in
\cite{Gogberashvili:2008xb}, but it is too complicated for our
purposes. In order to find something simpler, as executed in
\cite{Gomis:2008uv} we define a metric tensor $\gamma_{\mu\nu}$ from
the bilinear scalar product  Tr$(-,-)$, so  that 
\begin{equation}
\gamma_{\mu\nu}=\mbox{Tr}(E_\mu,\,E_\nu).\label{gamma}
\end{equation}
For the octonion case, $\gamma^{\mu\nu}=-\delta^{\mu\nu}$, and in the split-octonion case the
non-zero elements come from the diagonal in the multiplication table
and are 
\begin{equation}\gamma^{00}=\gamma^{11}=-\gamma^{22}=\gamma^{33}=\gamma^{44}=\gamma^{55}=-\gamma^{66}=-\gamma^{77}=1,\label{gamma_so}
\end{equation}
with $\gamma^{\mu\nu}=0$ to $\mu\neq\nu$ in both the cases. From the
structure constants we write,
\begin{equation}
c_{\mu\nu\kappa}E_\kappa=b_{\mu\nu}^{\hspace{3.5mm}\kappa}E_\kappa=b_{\mu\nu\lambda}\gamma^{\lambda\kappa}E_\kappa,
\end{equation}
\noindent where $b_{\mu\nu\kappa}$ is a totally anti-symmetric tensor that we call
the structure tensor, and $\gamma^{\mu\nu}$ works as a diagonal
metric tensor. The same letters for both the $\mathbb{O}$
and $\mathbb{SO}$ will be used, and will be differentiated explicitly in
the text where necessary. With this choice, for the octonionic case, we have simply
$b_{\mu\nu\kappa}=-c_{\mu\nu\kappa}$, and for the split-octonionic case,
\begin{equation}
b_{124}=b_{235}=b_{346}=b_{457}=b_{561}=-b_{672}=b_{713}=1.
\end{equation}
Now we can define a multiplication law which unifies the octonion and
the split-octonion cases, namely,
\begin{equation}
E_\mu E_\nu=\gamma_{\mu\nu}+b_{\mu\nu}^{\hspace{3.5mm}\kappa}E_\kappa,
\end{equation}
\noindent where, of course,
$b_{\mu\nu}^{\hspace{3.5mm}\kappa}=\gamma^{\kappa\lambda}b_{\mu\nu\lambda}$.
We also introduce a scalar product notation for the imaginary part of
the $\mathbb{(S)O}$, so that 
\begin{equation}
\gamma^{\mu\nu}A_\mu B_{\mu}=A\cdot B\qquad\mbox{and}\qquad \gamma^{\mu\nu}A_\mu E_{\mu}=A\cdot E.
\end{equation}
\noindent We adopt a notation so that octonions and split-octonions
are denoted by lower case Latin letters and their components by upper
case Latin letters. Thus, two generic elements, $a$ and $b$, are written as  $a=A_0+A\cdot E$
and $b=B_0+B\cdot E$, where the $E_0$ base component is superfluous,
and thus omited. We also introduce the notation where the contracted
indices of the structure tensor become equal to their components, or,
\begin{equation}
b_{\mu\nu}^{\hspace{3.5mm}\kappa}A_\kappa=b_{\mu\nu A}.\label{b_mnA}
\end{equation}
%
%
\noindent We stress that $A$ in (\ref{b_mnA}) has a very different
meaning from the indices $\mu$ and $\nu$. $A$ is a real valued
parameter and the indexes are discrete; we can also say that $A$, as
an index, has
neither the covariant nor the contravariant behaviors found for
$\mu$ and $\nu$. On the other hand, as the notation keeps the
anti-symmetry properties of the structure tensor $b_{\mu\nu\lambda}$,
it is a convenient way of doing the calculations. We describe several examples using
the formalism.
\subsection{Product of two elements}
Two elements $a=A_0+A\cdot E$ and $b=B_0+B\cdot E$ such that
$a,b\in\mathbb{(S)O}$  can be multiplied and the result is,
\begin{equation}
ab=A_0B_0+A \cdot B+ A_0(B\cdot E) + B_0 (A\cdot E) +b_{ABE},\label{ab}
\end{equation}
%
%
\noindent where, of course, $b_{ABE}=A_\mu B_\nu E_\kappa
b^{\mu\nu\kappa}=A_\mu B_\nu
E_\kappa\gamma^{\mu\mu^\prime}\gamma^{\nu\nu^\prime}\gamma^{\kappa\kappa^\prime}b_{\mu^{\prime}\nu^{\prime}\kappa^{\prime}}$
and $b_{ABE}$ mantains the anti-symmetry properties of the indices. In
this notation the covariant properties of the $b_3$ tensor are, of course, not
visible, as an example, $b_{ABE}=b^A_{\hspace{1,75mm}BE}$. 
\subsection{Properties of the structure tensor $b_{\mu\nu\lambda}$,}
The introduced notation is particularly useful to work with because the
properties already known to the octonion structure constants
can be rewritten for $b_{\mu\nu\lambda}$. As an analogy to the structure constant of
octonions, which have the dual tensor $C_{\mu\nu\lambda\kappa}$, we define 
the $B_{\mu\nu\lambda\kappa}$ tensor, dual to $b_{\mu\nu\lambda}$ and expressed as,
\begin{equation}
B_{\mu\nu\rho\sigma}=\frac{1}{6}\epsilon_{\mu\nu\rho\sigma\kappa\lambda\eta}b^{\kappa\lambda\eta}.
\end{equation}
\noindent For $\mathbb{O}$, we have the non-zero components of $B_{\mu\nu\kappa\lambda}$ as
\begin{equation}
B_{1275}=B_{1236}=B_{1435}=B_{1467}=B_{2473}=B_{2465}=B_{3657}=1,
\end{equation}
\noindent and for $\mathbb{SO}$ we have the same non-zero components as
$\mathbb{O}$, with the difference that all the values are given by
$B_{\mu\nu\kappa\lambda}=-1$, with the exception of $B_{1435}=1$. 
 The dual structure tensor $B_{\mu\nu\kappa\lambda}$ has many useful
properties. First, using the usual commutator $[-,-]$, we write the Jacobian, 
\begin{equation}
J_{\mu\nu\kappa}=[E_\mu,\,[E_\nu,\,E_\kappa]]+[E_\nu,\,[E_\kappa,\,E_\mu]]+[E_\kappa,\,[E_\mu,\,E_\nu]].\label{jacu}
\end{equation}
\noindent The Jacobian (\ref{jacu}) is identically zero for ordinary
associative Lie-algebras, but for $\mathbb{(S)O}$ it is related to $B_4$ as
\begin{equation}
J_{\mu\nu\kappa}=-3B_{\mu\nu\kappa
  E}=b_{[\nu\kappa}^{\hspace{3.5mm}\lambda}b_{\mu]\lambda E}\label{Jacobi}.
\end{equation}
\noindent In (\ref{Jacobi}), the indexes are anti-symmetrized with unit
weight, as explained in the appendix. Identities involving
the structure constants and the anti-symmetric $C_{\mu\nu\kappa\lambda}$ tensor, defined for the
octonion algebra \cite{Gunaydin:1995as}, can be adapted to the unified
formalism defined above and are summarized in the appendix.
\subsection{$\mathbb{(S)O}$  as Malcev algebras}
A Malcev algebra is defined through a commutator,
so that $[a,b]=\frac{1}{2}(ab-ba)$. It is known that $\mathbb{(S)O}$ are Malcev
algebras \cite{Carrion:2001pe}, and we verify it with the developed
formalism. For $x,y,z \in\mathbb{(S)O}$, the prescription to be satisfied is,
\begin{align}
&[x,x]=0\label{xx}\\
&J\left(x,y,[x,z]\right)=\left[J(x,y,z),x\right]\label{jaco}
\end{align}
%
%
%
\noindent where $J(x,y,z)$ is the Jacobian,
\begin{equation}
J(x,y,z)=[[x,y],z]+[[y,z],x]+[[z,x],y].
\end{equation}
\noindent Using (\ref{ab}) and the properties of the structure tensor,
we obtain
\begin{equation}
[[x,y],z]=b_{XY\mu}b^{\mu}_{\hspace{1.75mm}ZE}\qquad\mbox{and}\qquad J(x,y,z)=3 B_{XYZE},
\end{equation}
\noindent which allows us to write,
\begin{equation}
J\left(x,y,[x,z]\right)=-6\left[b_{XYZ}(X\cdot E)-b_{XYE}(X\cdot Z)-b_{YZE}(X\cdot X)-b_{ZXE}(X\cdot Y)\right]
\end{equation}
\noindent which satisfies both sides of (\ref{jaco}). As (\ref{xx}) is
trivially satisfied, so $\mathbb{(S)O}$ is a Malcev algebra. 
\section{Super-symmetric affinization\label{super-affin}}
We can use the notation indroduced above and the fact that
$\mathbb{SO}$ is a Malcev algebra to construct a super-current
algebra. This kind of structure is useful in string theory to
describe, for example, massive modes from a toroidal compactification \cite{Kaku:2000scf,Theisen:1989lnp}.
The algebra which describes the symmetry relating these modes
is an affine Lie algebra, which can be obtained by affining a Lie
algebra, or ever super-affining it in the case when there are fermions
involved. In the case we are dealing with, there is already  a
super-affining procedure executed by \cite{Carrion:2001pe} for
octonions, and it can be generalized for the split-octonion commutator
algebra as follows. 

For each generator, $g_\mu$, of a Lie algebra, $\mathfrak{g}$, with
structure constants, $f_{\mu\nu\lambda}$, we associate the fermionic super-field,
\begin{equation}
\Psi_\mu(X)=\psi_\mu(x)+\theta\phi(x),\label{PSI}
\end{equation}
\noindent where $X=(x,\,\theta)$ denotes the super-space with $\theta$
as a Grassmanian variable with $\theta^2=0$. $\psi(x)$ is a spin$-\frac{1}{2}$
fermionic field, and $\phi(x)$ a spin$-1$ bosonic field. The
super-affine $\hat{\mathfrak{g}}$ is introduced through, 
\begin{equation}
\{\Psi_\mu(X),\,\Psi_\nu(Y)\}=f_{\mu\nu\lambda}\,\Psi_\lambda(Y)\,\delta(X,Y)+\kappa\,\mbox{tr}(g_\mu,g_\nu)\,D_Y\delta(X,Y),\label{Psi_prod}
\end{equation}\noindent where $\kappa\in\mathbb{R}$, $\delta(X,Y)=(\theta-\eta)\delta(x-y)$, is a
super-symmetric delta function with $Y=(y,\eta)$ and
$D_Y=\partial_\eta+\eta\partial_y$ is a super-symmetric derivative. As
non-associative algebras are not necessarily represented by matrices
like Lie algebras \cite{Daboul:1999xv}, we must redefine the trace
that appears in (\ref{Psi_prod}). In the case where $g_\mu=E_\mu \in \mathbb{(S)O}$, we have,
\begin{equation}
\{\Psi_\mu(X),\,\Psi_\nu(Y)\}=f_{\mu\nu\lambda}\,\Psi_\lambda(Y)\,\delta(X,Y)+\kappa\,\Pi(E_\mu,E_\nu)\,D_Y\delta(X,Y),\label{PSIPSI}
\end{equation}\noindent where $\Pi(E_\mu,E_\nu)$ is a projection over the identity in
the composition law, in other words,
\begin{equation}
\Pi(E_\mu,E_\nu)=\mathfrak{Re}(E_\mu,E_\nu).
\end{equation}
\noindent So, with (\ref{PSI}) on the left hand side of (\ref{PSIPSI}) we get,
\begin{equation}
\{\Psi_\mu(X),\,\Psi_\nu(Y)\}=\{\psi_\mu(x),\,\psi_\nu(y)\}-\eta\,[\psi_\mu(x),\,\phi_\nu(y)]+\theta\,[\phi_\mu(x),\,\psi_\nu(y)]+\theta\eta\,[\phi_\mu(x),\,\phi_\nu(y)].
\end{equation}Comparing the above result with the substitution of
(\ref{PSI}) on the right hand side of (\ref{PSIPSI}) in terms of the
orders of $\eta$ and $\theta$ we obtain $\widehat{\mathcal{(S)O}}$ that is  
\begin{eqnarray}
&&\{\psi_\mu(x),\,\psi_\nu(y)\}=\kappa\,\delta_{\mu\nu}\delta(x-y)\\
&&[\psi_\mu(x),\,\phi_\nu(y)]=f_{\mu\nu\lambda}\,\psi_{\lambda}(y)\delta(x-y)\\
&&[\phi_\mu(x),\,\phi_\nu(y)]=\kappa\,\delta_{\mu\nu}\partial_y\delta(x-y)-f_{\mu\nu\lambda}\,\phi_{\lambda}(y)\delta(x-y)
\end{eqnarray}
\noindent In the above equation, $\psi_\mu(x)$ and   $\phi_\mu(x)$ are real fields and $\Psi_0(X)$ was associated with the $iE_0$ octonion.
The super-affine $\widehat{\mathcal{(S)O}}$ algebra is also a Malcev
super-algebra. Defining $\epsilon_x$ as $0$ or $1$ according to the
bosonic or fermionic character of $x\in \widehat{\mathcal{(S)O}}$ we have the graded bracket,
\begin{equation}
[x,\,y]=(-1)^{\epsilon_x\epsilon_y+1}[y,\,x].
\end{equation}
\noindent The super-Jacobian is,
\begin{equation}
J(x,y,z)=(-1)^{\epsilon_x\epsilon_z}[x,\,[y,\,z]]+(-1)^{\epsilon_z\epsilon_y}[z,\,[x,\,y]]+(-1)^{\epsilon_y\epsilon_x}[y,\,[z,\,x]],
\end{equation}

\noindent As $J(x,y,z)$ and $x$ satisfy (\ref{jaco}) and (\ref{xx}), we
have a Malcev algebra. So, we have characterized the super-symmetric
affinization of the (split-)octonion algebra  $\widehat{\mathcal{(S)O}}$, and shown that
the proposed formalism unifies both the octonion algebras in the same
formula.
\section{The generalized split Lie $3-$algebra\label{gsl3}}
\noindent Now, as executed by Yamazaki \cite{Yamazaki:2008gg}, who
constructed a realization of the generalized Lie $3-$algebra using
$\mathbb{O}$, we have to define a $3-$bracket. Accordingly, we
define left (L) and right (R) operators, $L,R:\mathbb{(S)O}\to\mathbb{(S)O}$ which work as, 

\begin{equation}
L_a\,b=ab\qquad\mbox{and}\qquad R_a\,b=ba
\end{equation}

\noindent We also define the derivative operator \cite{Nambu:1973qe},

\begin{eqnarray}
D_{a,b}x&=&\left([L_a,\,L_b]+[R_a,\,R_b]+[L_a,\,R_b]\right)x\nonumber\\
&=&\frac{1}{2}\left[a(bx)-b(ax)+(xb)a-(xa)b+a(xb)-(ax)b\right].\label{Dab}
\end{eqnarray}
\noindent Using the flexibility propriety of alternative algebras, such
as $\mathbb{(S)O}$, which says that the associator $(a,x,b)=(ax)b-a(bx)$ obeys,
\begin{equation}
(a,x,b)=-(b,x,a),
\end{equation}
\noindent we discover that (\ref{Dab}) is anti-symmetric in $a$ and $b$, namely
$D_{a,b}x=-D_{b,a}x$. Using 
\begin{equation}
b_{\kappa\mu\nu}\,b^{\kappa}_{\hspace{1.75mm}\rho\sigma}=B_{\mu\nu\rho\sigma}-\gamma_{\mu\rho}\gamma_{\nu\sigma}+\gamma_{\mu\sigma}\gamma_{\nu\rho}\qquad\mbox{and}\qquad
b_{\kappa\mu\nu}\,B^{\kappa}_{\hspace{1.75mm}\rho\eta\sigma}=3\left(b_{\mu\left[\rho\eta\right.}\,\gamma_{\left.\sigma\right]\nu}-b_{\nu\left[\rho\eta\right.}\,\gamma_{\left.\sigma\right]\mu}\right)
\end{equation}
\noindent one can write, for $a,b,x\in\mathbb{(S)O}$,
\begin{align}
[L_a,L_b]x&=\frac{1}{2}\left[a(bx)-b(ax)\right]\nonumber\\
&=b_{ABX}+X_0\;b_{ABE}-B_{ABXE}+(B\cdot X)(A\cdot E)-(A\cdot X)(B\cdot E)\\
[R_a,R_b]x&=\frac{1}{2}\left[(xb)a-(xa)b\right]\nonumber\\
&=b_{BAX}+X_0\;b_{BAE}-B_{ABXE}+(B\cdot X)(A\cdot E)-(A\cdot X)(B\cdot E)\\
[L_a,R_b]x&=\frac{1}{2}\left[a(xx)-(ax)b\right]\nonumber\\
&=B_{ABXE}
\end{align}
\noindent one can write (\ref{Dab}) as,
\begin{equation}
D_{a,b}x=2\left[(B\cdot X) (A\cdot E)-(A\cdot X)( B\cdot E) \right] - B_{ABXE},\label{D_comp}
\end{equation}
\noindent and using (\ref{D_comp}) we can prove that
\begin{equation}
D_{a,b}(xy)=(D_{a,b}x)y+x(D_{a,b})x\label{Deriv}.
\end{equation}
\noindent The property (\ref{Deriv}) is similar to the derivative
of a product property of real analysis, and so (\ref{Dab}) is known as
a derivative operator. This operator is used to define the $3-$bracket product,
\begin{equation}
[a,b,x]=D_{a,b}x,\label{3al}
\end{equation}
\noindent where $a,b,x\in\mathbb{(S)O}$. From (\ref{3al}) we wish to
construct a Lie $3-$algebra, and we thus adopt the following
definition \cite{Cherkis:2008qr}:
\paragraph*{\bf Definition\label{gl3a}}

{\it A generalized Lie $3-$algebra is an algebra $\mathbb{A}$ endowed with a
 $3-$product $[-,-,-]:\mathbb{A}^3\to\mathbb{A}$ and a bilinear
 positive product $(-,-)$ whose $a,b,c,x,y\in\mathbb{A}$ satisfy,}
\begin{enumerate}
\item Fundamental Identity
\begin{equation}
\left[x,y,[a,b,c]\right]=[[x,y,a],b,c]+[a,[x,y,b],]+[a,b,[x,y,c]]\label{fund}
\end{equation}
\item Metric Compatibility Condition 
\begin{equation}
([a,b,x],y)+(x,[a,b,y])=0
\end{equation}
\item Additional Symmetry Property
\begin{equation}
([x,y,a],b)-(a,[x,y,b])=0.
\end{equation}
\end{enumerate}
\noindent Yamazaki \cite{Yamazaki:2008gg} has proved that the above
definition is satisfied by $\mathbb{O}$, and our wish is to extend it
to $\mathbb{SO}$. The property (\ref{Deriv}) is enough to prove that the
fundamental identity of the definition is satisfied by
(\ref{3al}) in both cases. Now we define a bilinear product, so
that, for $a,b\in\mathbb{(S)O}$, we have,
\begin{align}
(a,b)&=\mbox{Re}(a,\bar{b})\\
&=\gamma^{\mu\nu} A_\mu B_\nu.
\end{align}
\noindent where Re$(a)$ picks the real part of $a\in\mathbb{(S)O}$ out
and $\bar{a}$ is the conjugate complex of $a$. The properties,
\begin{equation}
(ab,x)+(a,bx)=0\qquad\mbox{and}\qquad (ab,xy)-(ba,yx)=0
\end{equation}
\noindent are enough to demonstrate the Metric Compatibility Condition
and the Additional Symmetry Property of the definition. On the other
hand, the positive definiteness required for the bilinear product is
satisfied by $\mathbb{O}$, but not by $\mathbb{SO}$ , as this latter
case does not have a positive definite metric tensor. So we define a Split
Generalized Lie $3-$algebra as having the very definition above with
the only difference that $(-,-)$ is not positive definite.
\section{A possible gauge theory\label{gauge}}
Now we discuss several conjectures about a gauge theory based on the
split Lie $3-$algebra (\ref{3al}) and its symmetry group, which get its
structure constants from the algebra
\begin{equation}
[E_\mu,\,E_\nu,\,E_\lambda]=f_{\mu\nu\lambda}^{\hspace{5.25mm}\kappa}E_\kappa.
\end{equation}
Gomis {\it et al.} \cite{Gomis:2008uv} have discussed the conditions
the structure constants $f_{\mu\nu\kappa\lambda}$ have to satisfy in
order to construct a Lagrangian of a Bagger
and Lambert (BL) gauge theory. These conditions are simply the
fundamental identity (\ref{fund}) and  totally anti-symmetrical to the
indices. On the other hand Gustavsson \cite{Gustavsson:2008dy} has
briefly discussed the possibility of having 
non totally anti-symmetric structure constants. Yamazaki \cite{Yamazaki:2008gg}
has provided an example of such a theory \cite{Cherkis:2008qr} to the generalized
Lie $3-$algebra in the case of octonions, whose structure
constants were described as satisfying the relations
\begin{equation}
f_{\mu\nu\kappa\lambda}=-f_{\nu\mu\kappa\lambda}=-f_{\mu\nu\lambda\kappa}=f_{\kappa\lambda\mu\nu}.\label{f_part}
\end{equation}
\noindent As the derivation operator belongs to the group of
auto-morphisms of $\mathbb{O}$, which is known to be the special Lie
group $G_2$, this is naturally the gauge group of the theory. Some of
the properties of the structure constants components are,
\begin{align}
& f_{0\mu\nu\kappa}=0\\
& f_{124\mu}=f_{235\mu}=f_{346\mu}=f_{457\mu}=f_{561\mu}=f_{672\mu}=f_{713\mu}=0.
\end{align}
\noindent The above results show that these zero components come either from a
product that involves the $E_0$ component or from the seven
associative sub-algebras. The zero and non-zero components are 
the same both to octonion and split-octonion algebras. On the other
hand, the non-zero structure constants can be
decomposed into a totally anti-symmetric and a pairwise antisymmetric
part which satisfy (\ref{f_part}). Namely, we have,
\begin{equation}
f_{\mu\nu\kappa\lambda}=t_{\mu\nu\kappa\lambda}+\left(\delta_{\mu\kappa}\delta_{\nu\lambda}-\delta_{\mu\lambda}\delta_{\nu\kappa}\right)p_{\mu\nu\mu\nu},
\end{equation}
\noindent where $t_{\mu\nu\kappa\lambda}$ is  totally anti-symmetric
and $p_{\mu\nu\mu\nu}$ satisfies (\ref{f_part}). For the octonion case, we have, 
\begin{align}
&f_{1257}=-f_{1236}=-f_{2347}=f_{3415}=f_{4526}=f_{5637}=-f_{6712}=1\\
&f_{\mu\nu\mu\nu}=-2
\end{align}
\noindent and for the split-octonion case we have
\begin{align}
f_{1257}=-f_{1236}=&-f_{2347}=-f_{3415}=f_{4526}=f_{5637}=-f_{6712}=-1\\
f_{\mu\nu\mu\nu}=\;2,\qquad&\mbox{if}\qquad E_\mu^2\neq E_\nu^2\qquad \mbox{and}\\
f_{\mu\nu\mu\nu}\;=-2,\qquad&\mbox{if}\qquad E_\mu^2=E_\nu^2.
\end{align}
\noindent The non-zero components have some curious features. The $t_{\mu\nu\kappa\lambda}$
components correspond to the cosets of the associative
sub-algebras. For example, the coset of the $\{\pm E_0, \pm E_1, \pm E_2, \pm
 E_4\}$ subgroup is $\{\pm E_3, \pm E_5, \pm E_6, \pm E_7\}$, and we have
$f_{5637}$ as a non-zero component. The same occurs with the
sub-algebras generated by $\{E_0,\,E_\mu\}$.The cosets of these
sub-algebras generate the other non zero components of the structure
constants.

Far from curiosities, we can say that these results present a feature
of the structure constants of the algebra that has not been described
up until this point. Of course, the details about the structure constants are relevant to
the technical construction of the gauge theory. For the $\mathbb{SO}$ case,
the Lie $3-$algebra has the non-compact $G_2$  as its group of 
auto-morphisms \cite{Gogberashvili:2008xb, Beckers:1986vz, Beckers:1986dw}, and so
this is the gauge group of the theory. The $M2$-brane theories dual to
these gauge theories are not known, but if the conjecture of 
correspondence is correct, the split-octonion case is 
a realizations to the non-compact $G_2$ gauge group in the same sense
that $G_2$ is a realization of the octonion case. Studies in this
direction are currently being developed.
\section*{Appendix}
Here we give useful identities obeyed by the structure tensors $b_{\nu\nu\kappa}$
and $B_{\mu\nu\kappa\lambda}$. These relations were calculated based on former identities
involving the structure constants of octonions summarized by Gunayidin
and Ketov \cite{Gunaydin:1995as}. The square brackets denote an
anti-symmetrized product. As an example, the anti-symmetrized product of
$U_i$ and $V_{n-i}$ is given, by,
\begin{equation}
U_{\left[a_1\dots a_i\right.}V_{\left.a_{i+1}\dots
    a_n\right]}=\frac{1}{n!}\sum_{\sigma(a_1\dots
  a_n)}\mbox{sign}(\sigma)U_{\sigma\left(a_1\dots a_i\right.} V_{a_{i+1}\left.\dots a_n\right)}
\end{equation}
\noindent where $\sigma(a_1\dots a_n)$ gives all the $n!$ permutations
of the $n$ indexes and sign$(\sigma)$ gives a positive sign to an even
number of permutations and a negative sign to an odd number of permutations. Now we write the identities,
\begin{eqnarray}
&&b^2=-42\\
&&B^2=168\\
&&b_{\mu\kappa\lambda}\,b_{\nu}^{\hspace{1.75mm}\kappa\lambda}=-6\gamma_{\mu\nu}\\
&&b_{\rho\left[\mu\nu\right.}\,b_{\left.\lambda\right]\eta\sigma}+b_{\sigma\left[\mu\nu\right.}\,b_{\left.\lambda\right]\rho\eta}+b_{\eta\left[\mu\nu\right.}\,b_{\left.\lambda\right]\sigma\rho}=B_{\rho\eta\left[\mu\nu\right.}\gamma_{\left.\lambda\right]\sigma}+B_{\eta\sigma\left[\mu\nu\right.}\gamma_{\left.\lambda\right]\rho}+B_{\sigma\rho\left[\mu\nu\right.}\gamma_{\left.\lambda\right]\eta}-\nonumber\\
&&\hspace{6cm}-\gamma_{\eta\nu}\gamma_{\rho\sigma}\gamma_{\tau\mu}-\gamma_{\sigma\nu}\gamma_{\tau\rho}\gamma_{\eta\mu}-\gamma_{\tau\nu}\gamma_{\eta\rho}\gamma_{\sigma\mu}\\
&&b_{\kappa\mu\rho}\,b^{\kappa}_{\hspace{1.75mm}\nu\lambda}+b_{\kappa\mu\lambda}\,b^{\kappa}_{\hspace{1.75mm}\nu\rho}=\gamma_{\mu\rho}\gamma_{\nu\lambda}-\gamma_{\mu\lambda}\gamma_{\nu\rho}-2\gamma_{\mu\nu}\gamma_{\rho\lambda}\\
&&b_{\kappa\mu\nu}\,b^{\kappa}_{\hspace{1.75mm}\rho\sigma}=B_{\mu\nu\rho\sigma}-\gamma_{\mu\rho}\gamma_{\nu\sigma}+\gamma_{\mu\sigma}\gamma_{\nu\rho}\\
&&b_{\kappa\lambda\rho}\,B^{\kappa\lambda}_{\hspace{3.5mm}\mu\nu}=-4b_{\rho\mu\nu}\\
&&B_{\mu\kappa\lambda\eta}\,B_{\nu}^{\hspace{1.75mm}\kappa\lambda\eta}=24\gamma_{\mu\nu}\\
&&b_{\kappa\mu\rho}\,b^{\kappa}_{\hspace{1.75mm}\nu\sigma}=-\frac{1}{2}b_{\kappa\mu\nu}\,b^{\kappa}_{\hspace{1.75mm}\rho\sigma}+\frac{1}{2}B_{\mu\nu\rho\sigma}+\frac{1}{2}\left(\gamma_{\mu\rho}\gamma_{\nu\sigma}+\gamma_{\mu\sigma}\gamma_{\nu\rho}\right)-\gamma_{\mu\nu}\gamma_{\rho\sigma}\\
&&B_{\mu\nu\kappa\lambda}\,B^{\kappa\lambda}_{\hspace{3.5mm}\sigma\tau}=4\left(\gamma_{\mu\sigma}\gamma_{\nu\tau}-\gamma_{\mu\tau}\gamma_{\nu\sigma}\right)-2B_{\mu\nu\sigma\tau}\\
&&2b_{\kappa\left[\mu\nu\right.}\,B^{\kappa}_{\hspace{1.75mm}\left.\rho\eta\right]\sigma}=b_{\kappa\left[\mu\nu\right.}\,B^{\kappa}_{\hspace{1.75mm}\left.\rho\right]\eta\sigma}-b_{\kappa\eta\left[\mu\right.}\,B^{\kappa}_{\hspace{1.75mm}\left.\nu\rho\right]\sigma}\\
&&b_{\kappa\tau\left[\mu\right.}\,B^{\kappa}_{\hspace{1.75mm}\left.\nu\rho\right]\sigma}-b_{\kappa\sigma\left[\mu\right.}\,B^{\kappa}_{\hspace{1.75mm}\left.\nu\rho\right]\tau}=b_{\sigma\left[\mu\nu\right.}\,\gamma_{\left.\rho\right]\tau}-b_{\tau\left[\mu\nu\right.}\,\gamma_{\left.\rho\right]\sigma}\\
&&b_{\kappa\left[\mu\nu\right.}\,B^{\kappa}_{\hspace{1.75mm}\left.\sigma\right]\rho\eta}=-2\left(b_{\rho\left[\mu\nu\right.}\,\gamma_{\left.\sigma\right]\eta}-b_{\eta\left[\mu\nu\right.}\,\gamma_{\left.\sigma\right]\rho}\right)\\
&&b_{\kappa\mu\nu}\,B^{\kappa}_{\hspace{1.75mm}\rho\eta\sigma}=3\left(b_{\mu\left[\rho\eta\right.}\,\gamma_{\left.\sigma\right]\nu}-b_{\nu\left[\rho\eta\right.}\,\gamma_{\left.\sigma\right]\mu}\right)\\
&&b_{\kappa\left[\rho\eta\right.}\,B^{\kappa}_{\hspace{1.75mm}\left.\mu\nu\right]\tau}=2B_{\left[\mu\nu\rho\right.}\,\gamma_{\left.\eta\right]\tau}\\
&&b_{\left[\mu\nu\rho\right.}\,\gamma_{\left.\eta\right]\sigma}-b_{\left[\mu\nu\rho\right.}\,\gamma_{\left.\sigma\right]\eta}=\frac{3}{4}\left(b_{\sigma\left[\mu\nu\right.}\,\gamma_{\left.\rho\right]\eta}-b_{\eta\left[\mu\nu\right.}\,\gamma_{\left.\rho\right]\sigma}\right)\\
&&B_{\kappa\mu\nu\rho}\,B^{\kappa}_{\hspace{3.5mm}\eta\sigma\tau}=\gamma_{\eta\nu}\gamma_{\rho\sigma}\gamma_{\tau\mu}+\gamma_{\sigma\nu}\gamma_{\tau\rho}\gamma_{\eta\mu}+\gamma_{\tau\nu}\gamma_{\eta\rho}\gamma_{\sigma\mu}-\nonumber\\
&&\hspace{2.3cm}-3\left(B_{\eta\sigma\left[\mu\nu\right.}\,\gamma_{\left.\rho\right]\tau}+B_{\sigma\tau\left[\mu\nu\right.}\,\gamma_{\left.\rho\right]\eta}+B_{\tau\eta\left[\mu\nu\right.}\,\gamma_{\left.\rho\right]\sigma}\right)\\
&&B_{\kappa\sigma\left[\mu\nu\right.}\,B^{\kappa}_{\hspace{1.75mm}\left.\rho\eta\right]\tau}=b_{\sigma\left[\mu\nu\right.}\,b_{\left.\rho\eta\right]\tau}-B_{\mu\nu\rho\eta}\,\gamma_{\sigma\tau}-2\left(B_{\sigma\left[\mu\nu\rho\right.}\,\gamma_{\left.\eta\right]\tau}+B_{\tau\left[\mu\nu\rho\right.}\,\gamma_{\left.\eta\right]\sigma}\right)
\end{eqnarray}
{\bf Acknowledgments} Sergio Giardino thanks CNPq for its financial
support under grant number 152191/2008-9. 

%
%
%
%

\bibliographystyle{abbrv}
\makeatletter
\renewcommand\@biblabel[1]{#1.}
\makeatother 
\bibliography{bib_partida}

%
%
%
%

\end{document}